\documentclass[%
reprint,
superscriptaddress,
amsmath,amssymb,
aps,
]{revtex4-2}

\usepackage{graphicx}
\usepackage{dcolumn}
\usepackage{bm}
\usepackage{subfig}
\usepackage{color}
\usepackage[justification=raggedright,singlelinecheck=false]{caption}
\usepackage{lipsum}
\usepackage{soul}
\usepackage{xcolor}

\begin{document}
	\setstcolor{red}
	
	\title{Fast multi-qubit global-entangling gates without individual addressing of trapped ions}
	
	
	\author{Kaizhao Wang}
	\thanks{These two authors contributed equally to the work}
	\affiliation{
		State Key Laboratory of Low Dimensional Quantum Physics, Department of Physics, Tsinghua University, Beijing 100084, China
	}
	
	\author{Jing-Fan Yu}
	\thanks{These two authors contributed equally to the work}
	\affiliation{
		Institute for Interdisciplinary Information Sciences, Tsinghua University, Beijing 100084, P. R. China
	}
	
	\author{Pengfei Wang}
	\affiliation{
		State Key Laboratory of Low Dimensional Quantum Physics, Department of Physics, Tsinghua University, Beijing 100084, China
	}
	\affiliation{
		Beijing Academy of Quantum Information Sciences, Beijing 100193, China
	}
	\author{Chunyang Luan}
	\affiliation{
		State Key Laboratory of Low Dimensional Quantum Physics, Department of Physics, Tsinghua University, Beijing 100084, China
	}
	\author{Jing-Ning Zhang}
	\affiliation{
		Beijing Academy of Quantum Information Sciences, Beijing 100193, China
	}
	
	\author{Kihwan Kim}%
	\affiliation{
		State Key Laboratory of Low Dimensional Quantum Physics, Department of Physics, Tsinghua University, Beijing 100084, China
	}
	\affiliation{
		Beijing Academy of Quantum Information Sciences, Beijing 100193, China
	}
	\affiliation{
		Frontier Science Center for Quantum Information, Beijing 100084, People’s Republic of China
	}
	
	\date{\today}
	
	\begin{abstract}
		We propose and study ways speeding up of the entangling operations in the trapped ions system with high fidelity. First, we find a scheme to increase the speed of a two-qubit gate without the limitation of trap frequency, which was considered as the fundamental limit. Second, we study the fast gate scheme for entangling more than two qubits simultaneously. We apply the method of applying multiple frequency components on laser beams for the gate operations. In particular, in order to avoid infinite terms from the coupling to carrier transition, we focus on the phase-insensitive gate scheme here. We carefully study the effect of large excitation of motional mode beyond the limit of Lamb-Dicke approximation by including up to second order terms of the Lamb-Dicke parameter. We study the speed limit of multi-qubit global entangling gates without individual addressing requirements. Furthermore, our gates can be made insensitive to the fluctuation of initial motional phases which are difficult to stabilise in the phase-insensitive gate scheme.
	\end{abstract}
	
	\maketitle
	
	\section{\label{sec:Intro}Introduction\protect\\} Quantum computers can provide solutions for certain complex problems intractable by classical computers. The trapped-ion system is one of the most promising physical platforms to realize a practical quantum computer~\cite{Ladd2010-eo, Monroe2013-kh, Haffner2008-yf, Blatt2012-so}. The longest coherence time of single-qubits~\cite{Wang2017Single-qubit,Wang2021Single} and the highest fidelity of the entangling gates~\cite{Ballance2016-cx, Gaebler2016-wl,Clark2021-jl} among all physical platforms have been demonstrated in trapped-ion systems. Due to the long coherence time of a trapped-ion qubit, the number of gates can be implemented before decoherence can be in the order of 10$^8$, which is also in the leading position among other physical platforms for quantum computers. However, the speeds of entangling gates in the trapped-ion system can be considered relatively slow, which would limit the application of some quantum algorithms for practical purposes. There have been many theoretical proposals and studies to speed up the trapped-ion entangling gates~\cite{Garcia-Ripoll2005-ts, Garcia-Ripoll2003-kw, Duan2004-lr, Steane_2014, Torrontegui2020-xd, Mehdi2021-cw}. Recently, experimental realizations of such fast quantum gates have also been demonstrated with trapped ions~\cite{Schafer2018-xj, Wong-Campos2017-ot, Zhang2020-um}. Apart from speeding up the gate, by taking advantage of the full-connectivity of trapped ion system~\cite{Choi2014Optimal,debnath2016demonstration,Linke2017-dq}, simultaneous operations on more than two qubits~\cite{Figgatt2019-bj, Lu2019-yg, Shapira2020-xe, Grzesiak2020-jb} such as parallel gates and global gates have been intensively studied to shorten the running time of quantum algorithms. 
	
	The quantum entangling gates with trapped ion-qubits are primarily performed through the coupling of vibrational modes, where the speed of the gates is mainly limited by the frequencies of the relevant vibrational modes. The duration of two-qubit gates has been pushed to $1.6~\mu s$, which is around 3 times the period of the axial center of mass mode, without serious loss in fidelity~\cite{Schafer2018-xj}. A faster gate of 480 ns, slightly less than one motional period, has also been demonstrated, but the fidelity is much lower owing to the high excitation of vibrational modes out of the Lamb-Dicke regime~\cite{Steane_2014,Schafer2018-xj}.
	This out of Lamb-Dicke approximation error has been considered as a bottleneck to push the gate time below single vibrational period. As for multiple-qubit parallel or global gates, unlike the case of two-qubit gate, the speedup of which have not yet been thoroughly explored \cite{Figgatt2019-bj,Lu2019-yg}. 
	
	In this work, we demonstrate a way to significantly reduce the error from the Lamb-Dicke approximation breaking down by taking the higher order of Lamb-Dicke approximation into account \cite{Steane_2014,Schafer2018-xj}. We found various numerical solutions of two-qubit gates below one motional period with high fidelities. We also present a concrete example of two-qubit gate and the error budget for the gate which is in the order of 10$^{-3}$. We study the limits on the gate duration for more than two-qubit global entangling gates and compare them with equivalent circuits using  sequences of two-qubit gates. We found N-qubit global entanglement gates with duration comparable to those decomposed into multiple two-qubit gates of one vibrational period duration without requiring individual addressing. Our numerical methods are based on the multi-frequency method proposed by Ref.\cite{Shapira2020-xe}.

	
	\section{\label{sec:Theory}Theory\protect\\}

	\subsection{\label{subsec: GM} Theoretical framework for entangling gates in a linear ion-chain}
	
	\begin{figure}[htp!]
		\centering
		\includegraphics[width=1.0\columnwidth]{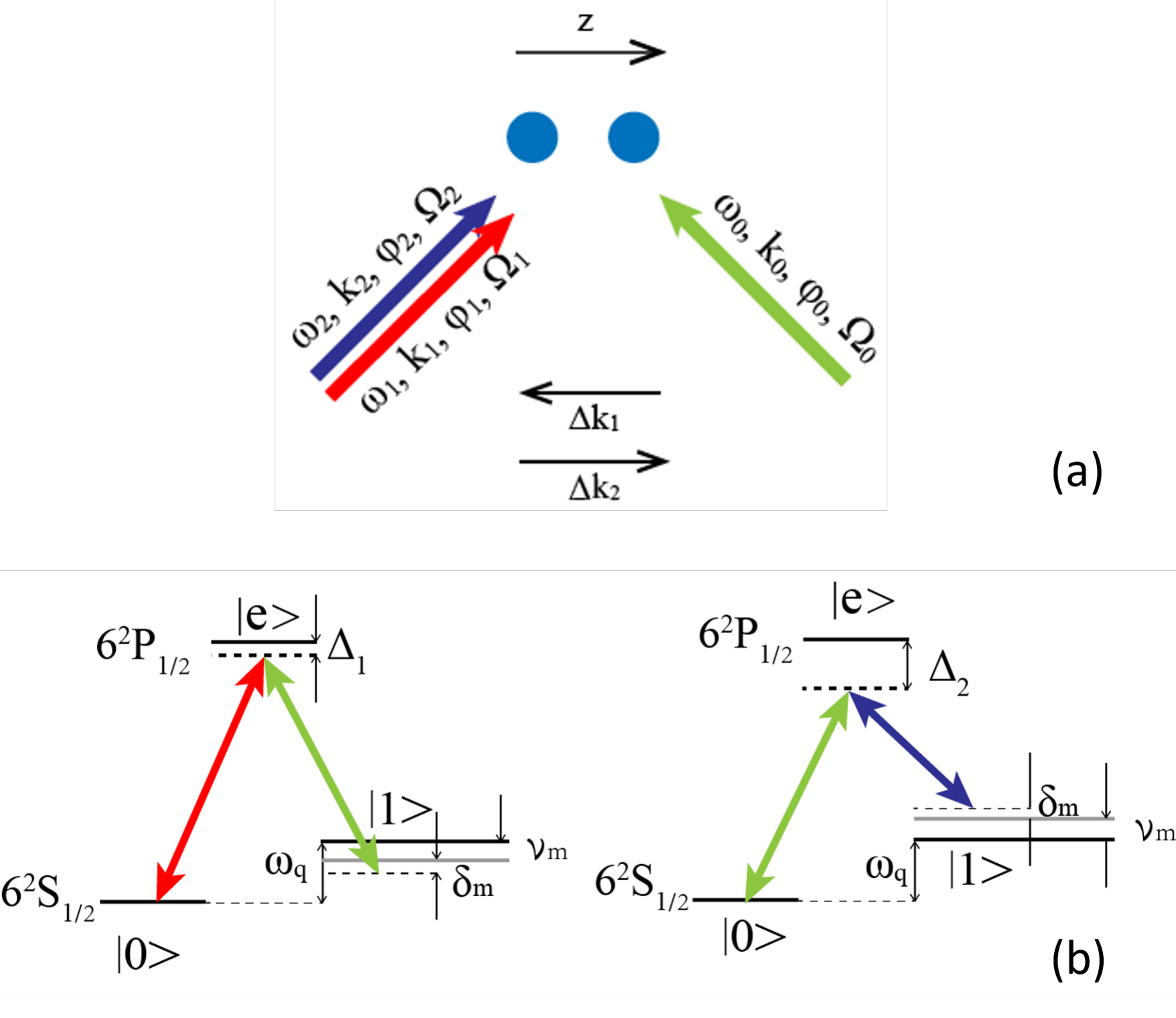}
		\caption{Configuration for the phase-insensitive M\o lmer-S\o renson gate. (a) One monochromatic and one bi-chromatic laser beams are applied on the target ions in the chain. Here $\omega_i,\;{\mathbf k}_i,\;\phi_i,$ and $\Omega_i$ represent the angular frequency, wave-vector, phase, and Rabi-frequency of the three laser beams respectively, where $i=0,1,2$. The effective wave-vectors of the two Raman transitions, $\Delta{\mathbf k}_1$ and $\Delta {\mathbf k}_2$, are in opposite directions. (b) The energy levels and couplings of the Raman laser beams for the gate process as an example of using $^{171}\mathrm{Yb}^+$ ions, where $|0\rangle ,\;|1\rangle$, and $|e\rangle$ represent the qubit levels and the intermediate level used in Raman transitions. $\Delta_1$ and $\Delta_2$ are the detuning from the intermediate level, and $\omega_q$ represents the qubit level splitting. $\nu_m$ is the motional frequency of the $m^{th}$ mode, and $\delta_m$ is the detuning from it.} \label{fig:configuration}
	\end{figure}
	
	For trapped-ion quantum computation, qubit states are encoded in internal energy levels of the ions which are typically trapped in linear Paul traps. With properly controlled laser beams, we first entangle qubit states with collective vibrational modes and use these collective motions as information bus to entangle different qubits. At the end of the gate operation, we disentangle the qubit and motional modes. Two main types of entangling gates used with trapped ions are $\hat\sigma_{\phi_S}\hat\sigma_{\phi_S}$ gate and $\hat\sigma_{z}\hat\sigma_{z}$ gate, where $\hat\sigma_{\phi_S}=\cos{\phi_S}~\hat\sigma_{x}+\sin{\phi_S}~\hat\sigma_{y}$, and $\hat\sigma_{x}$, $\hat\sigma_{y}$, and $\hat\sigma_{z}$ are the Pauli operators ~\cite{Soerensen1998-pz,Sorensen2000-if,haljan2005spin,Lee2005-lx}. With Raman laser beams for hyperfine qubits, $\hat\sigma_{\phi_S}\hat\sigma_{\phi_S}$ gates can be implemented in a phase-sensitive or phase-insensitive manner~\cite{haljan2005spin,Lee2005-lx}. We mainly focus on the phase-insensitive gate, which is formally equivalent to the $\hat\sigma_{z}\hat\sigma_{z}$ gate~\cite{haljan2005spin,Lee2005-lx}. In these gate schemes, we can take advantage of being able to derive the evolution in a mathematically neat way including the off-resonant carrier coupling terms \cite{Wu2018Noise}.
	
	To implement a phase-insensitive $\hat\sigma_{\phi_S}\hat\sigma_{\phi_S}$ gate, we shine two counter-propagating beams with equal but opposite detuning on the ions as shown in Fig.~\ref{fig:configuration}. Here we set $\phi_S =0$, which leads to $\hat\sigma_{\phi_s}\hat\sigma_{\phi_s}=\hat\sigma_{x}\hat\sigma_{x}$. In Fig.~\ref{fig:configuration}, we show the laser schemes for qubits encoded in sub-levels of the ground manifold. The blue and red detuned transitions are realized with $\Lambda$-type Raman transitions. The Hamiltonian of this system in the interaction picture can be written as
	\begin{eqnarray}\label{Eq:Hamiltonian_full}
		\hat{\mathcal H}=\sum_{j=1}^{N}\frac{\hbar\Omega}{2}\left[e^{i(\mu t+\phi_b+\vec{k}_b\cdot\hat{\vec{r}}_j)} + e^{i(-\mu t+\phi_r+\vec{k}_r\cdot\hat{\vec{r}}_j)} \right]\hat\sigma_{+}^j+h.c.,\nonumber\\
	\end{eqnarray}
	where $j$ indicates the $j^{th}$ ion. We assume that different ions are experiencing same effective Rabi frequency $\Omega$, phase $\phi_b$, $\phi_r$, and detuning $\mu$ from the carrier frequency. As the detuning is small compared to the optical frequency, we have $\vec{k}=\vec{k}_b\approx-\vec{k}_r$ and $\vec{k}\cdot\hat{\vec{r}}_j$ can be written as
	\begin{equation}
		\label{Eq:1}
		\vec{k}\cdot\hat{\vec{r}}_j=\sum_{m=0}^{N-1}\eta_{j,m}\left(\hat{a}_m^{\dagger}e^{i\nu_mt}+\hat{a}_me^{-i\nu_mt}\right),
	\end{equation}
	where $\hat{a}_m$ and $\hat{a}_m^{\dagger}$ are the annihilation and creation operators of the $m^{th}$ motional mode, respectively, and $\nu_m$ are the corresponding frequencies with $\eta_{j,m}=b_{j,m}\eta_{m}$, where $b_{j,m}$ is the $j^{th}$ entry of the eigen-vector of $m^{th}$ normal mode and $\eta_m=\Delta k\sqrt{\frac{\hbar}{2M\nu_m}}$ is the Lamb-Dicke parameter of $m^{th}$ mode. As typical choice of the Lamb-Dicke parameter($\eta_{m}$) is small, in the slow gate regime, we can neglect the $o(\eta)$ terms in the Hamiltonian. However, in the fast gate regime because of relatively large Rabi frequency and high excitation of the motional modes, we need to include higher orders. Up to the second-order terms, the original full Hamiltonian  (\ref{Eq:Hamiltonian_full}) can be written as,
	\begin{eqnarray}\label{Eq:Hsum}
		\hat{\mathcal{H}}&=&\hat{\mathcal{H}_0}+\hat{\mathcal{H}_1}+\hat{\mathcal{H}_2}+o(\eta^2), \nonumber
	\end{eqnarray}
	where
	\begin{eqnarray} 
		\hat{\mathcal{H}_0}&=&\sum_{j=1}^{N}\hbar\Omega\cos(\mu t+\phi)\hat\sigma_{x}^j \label{Eq:Hamiltonian0} \\\hat{\mathcal{H}_1}&=&-\sum_{j=1}^{N}\hbar\Omega\sin(\mu t+\phi)\left(\vec{k}\cdot\hat{\vec{r}}_j\right)\hat\sigma_{x}^j \label{Eq:Hamiltonian1}
		\\\hat{\mathcal{H}_2}&=&\frac{1}{2}\sum_{j=1}^{N}\hbar\Omega\cos(\mu t+\phi)\left(\vec{k}\cdot\hat{\vec{r}}_j\right)^2\sigma_{x}^j.\label{Eq:Hamiltonian2}
	\end{eqnarray}
	Here, $\phi=\frac{\phi_b-\phi_r}{2}$ is the phase difference between the blue-detuned and red-detuned beam. 
	
	Applying the Magnus expansion, we can obtain the time evolution of the gate. Here, we only remain the lowest order term in the Magnus expansion arising from $\hat{\mathcal{H}_2}$ in Eq.(\ref{Eq:Hamiltonian1}). Then, the evolution operator can be written as 
	\begin{widetext}
		\begin{equation}
			\hat U(\tau)\approx\exp\left\{-\frac{i}{\hbar}\int_{0}^{\tau_g}\left(\hat{\mathcal{H}}_0(t)+\hat{\mathcal{H}}_1(t)+\hat{\mathcal{H}}_2(t)\right)\mathrm{d}t-\frac{1}{2\hbar^2}\int_{0}^{\tau_g}\mathrm{d}t_1\int_{0}^{t_1}\mathrm{d}t\left[\hat{\mathcal{H}}_1(t_1),\;\hat{\mathcal{H}}_1(t)\right]\right\}, 
			\label{Eq:Evol}
		\end{equation}
	\end{widetext}
	
	As shown explicitly in Appendix~\ref{App: evol}, the first three integrals lead to single qubit rotation, phase space displacement and some second-order effect of the motional modes depending on the qubit states. The integral of the commutator generate non-trivial geometric phases that are dependant on the qubit-states. The evolution is thereby governed by four sets of parameters the single-qubit rotation angle $\Phi$, the displacement $\alpha_m$, the second-order effects coefficients $\beta_{m_1,\pm m_2}$ and the geometric phase $\Theta_{ij}$.
	
	The goal here is to find proper gate sequence such that at the end of the gate the following constraints are satisfied.
	\begin{eqnarray}
		\Phi(\tau_g)&{}=&0\label{Eq:cons1}\\
		\alpha_{m}(\tau_g)&{}=&0\quad \forall m\label{Eq:cons2}\\
		\beta_{m_1,\pm m_2}(\tau_g)&{}=&0\quad \forall m_1, m_2\label{Eq:cons3}\\
		\Theta_{ij}(\tau_g)&{}=&\frac{\pi}{4}\quad \forall i,j\label{Eq:cons4}.
	\end{eqnarray}
	Here, for ultra-fast gate, with gate time below or around one vibrational period, the second order term,  $\left(\vec{k}\cdot\hat{\vec{r}}_j\right)^2$ in Eq.(\ref{Eq:Hamiltonian1}) would lead to a significant error. We include the constraint, Eq.(\ref{Eq:cons3}) to address such out of Lamb-Dicke error.  
	
	In the paper, we consider three different ways of including those constraints as follows. For the first case, Type I, we include the constraints of Eq.(\ref{Eq:cons1}), Eq.(\ref{Eq:cons2}), and Eq.(\ref{Eq:cons4}) without Eq.(\ref{Eq:cons3}). Here, we assume that the initial phase $\phi$ can be controlled and stabilised in experiment. The phase $\phi$ is related to the optical phases of Raman laser beams, which is difficult to achieve \cite{delaubenfels2015modulating,schmiegelow2016phase}. We consider the Type II gate, where, apart from the constraints of Type I gate, additional constraints are included in order to make the gate robust against the fluctuation of the initial phase $\phi$. The detailed derivation and expression of the constraints for initial phase robustness are given in Appendix~\ref{App: Initial phase robustness condition}. The Type II gates are robust against initial phase fluctuation, but the performance of the gate is limited because of phonon excitation out of Lamb-Dicke regime during the gate operation. Therefore, for the Type III gate, we further include the second order constraints given by Eq.(\ref{Eq:cons3}) to suppress this higher-order effect.
	
	
	\subsection{\label{subsec: AM} Multi-frequency formalism}
	
	We numerically investigate the fast gates based on the method shown in Ref.\cite{Shapira2020-xe}. The method can be understood as the generalization of phase and intensity modulations of Raman laser beams by applying multi frequency components from the Fourier decomposition of the beams, which provides an efficient and systematic way of numerical gate sequence-searching. 
	
	Assuming the gate duration is $\tau_g$, we add pairs of blue-sideband and red-sideband frequencies with detuning $\pm\omega_n=\pm2 \pi n/\tau_{g}$ from the carrier frequency and phase difference $2\phi_n$. 
	The Hamiltonian with multi-frequency components can be obtained by replacing the $\Omega\cos(\mu t +\phi)$ in Eq.(\ref{Eq:Hamiltonian0}) and (\ref{Eq:Hamiltonian2}) and $\Omega\sin(\mu t+\phi)$ in Eq.(\ref{Eq:Hamiltonian1}) by the following two terms.
	\begin{eqnarray}
		f_{car}&=&\Omega\sum_{n\neq0}\cos(\omega_n t+\phi_n)\nonumber\\
		&=&\Omega\sum_{n\neq0}\left[c_n\cos(\omega_n t)-s_n\sin(\omega_n t)\right]\label{Eq:fcar}\\
		f_{sdf}&=&\Omega\sum_{n\neq0}\sin(\omega_n t+\phi_n)\nonumber\\
		&=&\Omega\sum_{n\neq0}\left[c_n\sin(\omega_n t)+s_n\cos(\omega_n t)\right]\label{Eq:fsdf}
	\end{eqnarray}
	where the $c_n$ and $s_n$ terms corresponding to setting $\phi_n$ to 0 and $\frac{\pi}{2}$ respectively.
	In the actual calculation we only consider finite frequency components, so we truncate the summation of $n$ to $n_e=\lfloor \frac{\mathrm{Max}(\nu_{m})}{2\pi}t_g\rfloor+15$. We define the vector $\vec{c}$ and $\vec{s}$ as
	\begin{eqnarray*}
		&\vec{c}=\begin{bmatrix}
			c_1\\
			\vdots\\
			c_{n_e}\\
			c_{-1}\\
			\vdots\\
			c_{-n_e}
		\end{bmatrix};\quad
		\vec{s}=
		\begin{bmatrix}
			s_1\\
			\vdots\\
			s_{n_e}\\
			s_{-1}\\
			\vdots\\
			s_{-n_e}.
		\end{bmatrix}
	\end{eqnarray*}
	
	The constraints given in Eq.(\ref{Eq:cons1})-Eq.(\ref{Eq:cons4}) can now be written in a matrix form. It is noticeable that the first condition, $\Phi=0$ is fulfilled automatically by setting $\omega_n$ as the integer multiple of $2\pi/\tau_{g}$. The other constraints related to $\alpha_m$, $\beta_{m_1, \pm_2}$ and $\Theta_m$ can be written as the following form.
	\begin{eqnarray}
		\alpha_m&=&\vec{a}_{m}^{T}\vec{\Omega}\label{eq:alpha},\label{Eq:alpha}\\
		\beta_{m_1, \pm m_2}&=&\vec{b}_{m_1, \pm m_2}^{\;T}\vec{\Omega}\label{eq:2ndcoeff}\label{Eq:beta}\\
		\Theta_{m}&=&\vec{\Omega}^{T}\tilde{\theta}_m\vec{\Omega}\label{eq:Theta}.
	\end{eqnarray}
	where $\vec{\Omega}=\Omega\begin{bmatrix}
		\vec{c}\\
		\vec{s}
	\end{bmatrix}$, and the explicit expression of $\vec{a}_{m}$, $\vec{b}_{m_1, \pm m_2}$ and $\tilde{\theta}_m$ are given in Appendix~\ref{App: Matrices expression}.
	
	If the initial phase $\phi$ can be stabilised, the coefficient before the even order terms of $\eta$ in the Hamiltonian (\ref{Eq:Hamiltonian_full}) can be set to 0 by setting $c_n=-c_{-n}$ and $s_n=s_{-n}$. The details of the derivation are discussed in the Appendix~\ref{App: Initial phase robustness condition}. Although it is possible to stabilise this phase \cite{delaubenfels2015modulating,schmiegelow2016phase}, which comes from optical phases of Raman laser beams, it is favorable to have the gate insensitive to fluctuation of the initial phase. This insensitivity to initial phase can be achieved by including additional terms from different initial phase by $\pi/2$ as $\vec{\Omega}'=\Omega\begin{bmatrix}
		-\vec{s}\\
		\vec{c}
	\end{bmatrix}$ and requiring both $\vec{\Omega}$ and $\vec{\Omega}'$ to fulfill the constraints in Eq.(\ref{eq:alpha})-Eq.(\ref{eq:Theta}). Detailed derivation and proof for the initial phase robustness can be seen in Appendix~\ref{App: Initial phase robustness condition}.
	
	
	In the numerical search, we first find the null space of the linear constraints and solve the quadratic equations. Denoting $L$ as the coefficient matrix of all the linear constraints given by  Eq.(\ref{Eq:alpha}) for Type I and Type II gates, and by both Eq.(\ref{Eq:alpha}) and Eq.(\ref{Eq:beta}) for Type III gate. We find out a basis of the kernel of $L$ by numerically solving the linear equations of $L\vec{\Omega}=0$. We define $K=(\vec{\Omega}_1,\; \vec{\Omega}_2,\; ...,\;\vec{\Omega}_l)$ as a basis of the kernel of $L$, $\mathrm{Ker}(L)$, where $l$ denotes the number of independent vectors in $\mathrm{Ker}(L)$. Therefore, $K$ is described by a $4n_e\times l$ matrix. Any vector in $\mathrm{Ker}(L)$ can then be represented as a linear combination of the basis vectors in $K$.
	\begin{equation}
		\vec{\Omega}=K\vec{x},\quad \forall\vec{\Omega}\in\mathrm{Ker}(L),
	\end{equation}
	where $\vec{x}$ is a $l$-dimensional real vector. We can rewrite Eq.(\ref{eq:Theta}) as
	\begin{equation}
		\Theta_m=\vec{x}^{T}K^{T}\tilde{\theta}_{m}K\vec{x}=\vec{x}^{T}A_{m}\vec{x},\label{eq:Theta_m}
	\end{equation}
	where $A_{m}=K^{T}\tilde{\theta}_{m}K$. Now, we just need to find a $l$-dimensional real vector $\vec{x}$ such that the quadratic form $\vec{x}^{T}A_m\vec{x}$ equals to our desired entangling phase. We numerically find the vectors $\vec{x}$ by minimizing the Rabi frequency $\Omega$ subject to the quadratic constraints set by Eq.(\ref{eq:Theta}). We minimize the Rabi frequency to make demanded laser power favorable for the experiment.
	
	
	

	\section{\label{sec:Results}Results\protect\\}
	
	\subsection{\label{subsec:two-qubit}Two-qubit gate}
	In the numerical calculation below, we use $^{171}\mathrm{Yb}^+$ ion, and $\Delta k$ is calculated assuming the two Raman beams from 355 nm laser are applied perpendicularly as shown in Fig.~\ref{fig:configuration}(a), thereby taking $\eta\approx0.136$ after normalizing the center-of-mass mode frequency $\nu_0=2\pi\times 1$ MHz in the axial direction, which corresponding 1 $\mu$s vibrational period. 
	
	\begin{figure}[htp!]
		\centering
		\includegraphics[width=1.0\columnwidth]{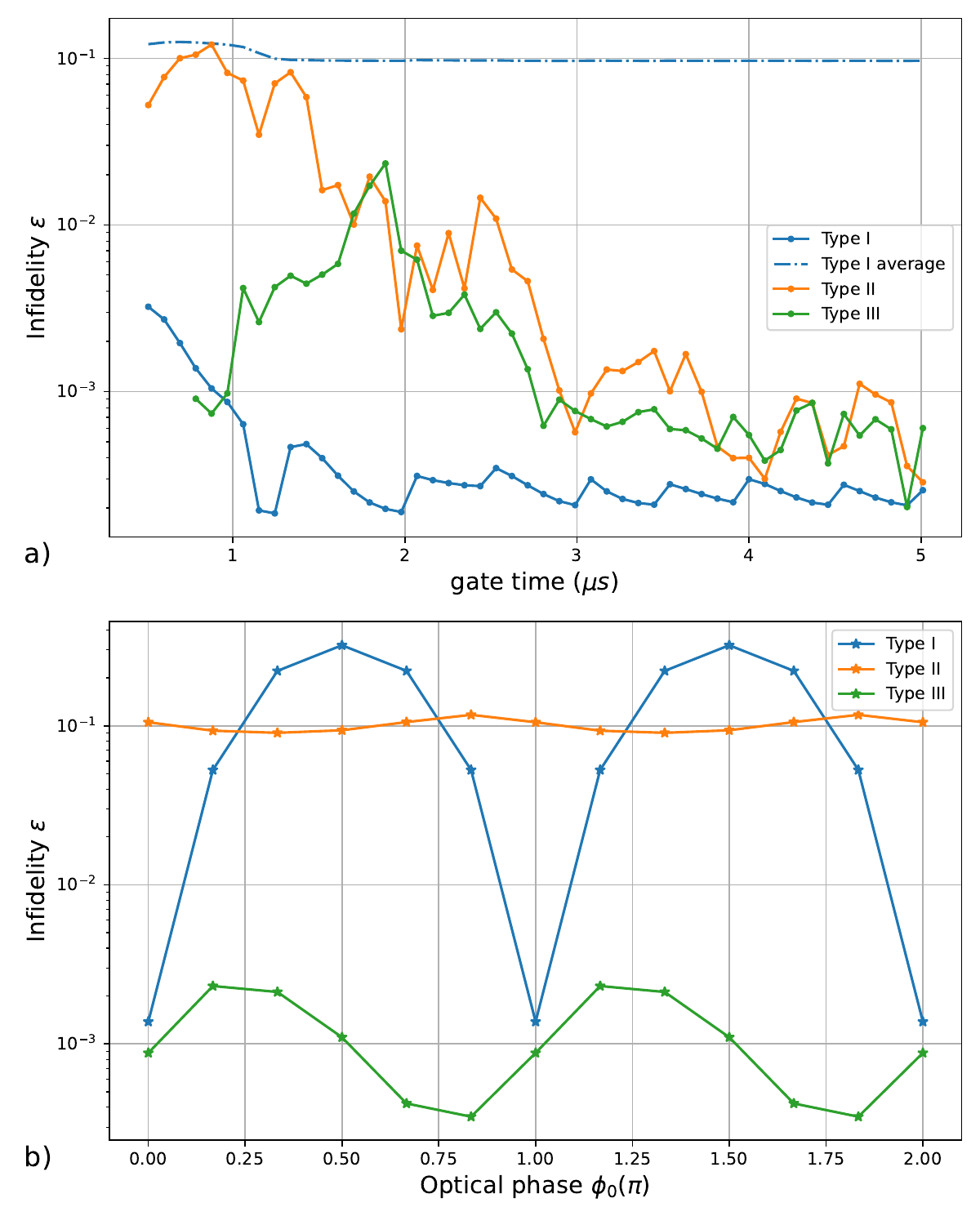}
		\caption{ (a) Infidelity of the fast two-qubit gate depending on the duration. Here one vibrational period of the axial center of mass mode is 1 $\mu$s, since the frequency is ($2\pi) \times$1 MHz. Therefore,  The infidelity is estimated from the time-evolution of the Hamiltonian (\ref{Eq:Hamiltonian_full}) without any approximation. The blue dots show the performance of the Type-I gates that require optical phase stabilization. The orange dots show the performance of the Type-II gates that are robust against initial phase fluctuations. The green dots show the performance of the Type-III gates that contain initial phase robustness and the 2nd order constraints given by Eq.(\ref{eq:2ndcoeff}). (b) Infidelity depending on the initial optical phases $\phi_0$ for the two-qubit gate with the duration of $\tau_g\approx0.79~\mu s$.}
		\label{fig:t-err}
	\end{figure} 
	
	
	We have applied the method discussed in~\ref{subsec: AM} to find solutions for ultra-fast two-qubit gates as shown in Fig.~\ref{fig:t-err}(a). We estimate the infidelity of the gates from the time evolution of the original Hamiltonian (\ref{Eq:Hamiltonian_full}) without any approximation. First, we study the Type-I gates, shown as blue dots in Fig.~\ref{fig:t-err}(a), that require the precise control and stabilisation of the initial phase $\phi$ shown in Eqs.(\ref{Eq:Hamiltonian0}-\ref{Eq:Hamiltonian2}), which comes from optical phases of Raman laser beams. When the initial phase are set to $\phi_0 = 0$, all the even order terms of $\eta$ in the original Hamiltonian (\ref{Eq:Hamiltonian_full}) can be removed by setting $c_n=-c_{-n}$ and $s_n=s_{-n}$ for all $n$. Here the effect of the second order in out-of-Lamb-Dicke approximation is suppressed. We observe decent performance of the gate with the duration below 1$\mu$s. For example, for the duration of $0.79~\mu$s, the infidelity is estimated as $1.3\times10^{-3}$. The details of the gate at the right phase $\phi_0 = 0$ are described in Appendix~\ref{App: 2-qubit example}. If the durations are between 1 and 2 $\mu$s, the infidelities are below $5\times10^{-4}$. However, if The optical phase is not stabilized \cite{schmiegelow2016phase,palmero2017fast}, the performance of the gate would be seriously degraded as shown in blue dashed dots of Fig.~\ref{fig:t-err}(a). The average infidelity over initial phases is larger than 10$\%$ for any gate duration. The example of the initial phase dependence for $0.79~\mu$s gate is shown in Fig.~\ref{fig:t-err}(b).   

	We investigate the Type II gates that are obtained with the robustness constraints against the optical phase fluctuations,  discussed in detail in Appendix~\ref{App: Initial phase robustness condition}. The infidelities of the Type II gates depending on the duration are as shown as orange dots in Fig.~\ref{fig:t-err}. As the duration approaches one vibrational period, the infidelities increase from $10^{-3}$ (larger than 3 $\mu$s) to $10^{-1}$ (around 1 $\mu$s), where the main source of the infidelity comes from out-of-Lamb-Dicke error. If we estimate infidelity of the Type II gate with the Hamiltonian with only the first-order Lamb-Dicke approximation (Hamiltonian (\ref{Eq:Hamiltonian0}) and (\ref{Eq:Hamiltonian1}), there exists negligible infidelity. Here the infidelities are from averaging over different initial optical phases. As shown in the orange dots of Fig~\ref{fig:t-err}(b) for the case of $\approx 0.79\mu s$, the dependence of initial phases are much more suppressed than that of the Type I gate. There still exists the $3\%$ dependence, which is mainly because the phase dependence in the higher order is not considered. The details of the gate are described in Appendix~\ref{App: 2-qubit example}.
	
	We mitigate this problem by taking into account the second-order terms in the constraints as shown in Eq.(\ref{Eq:cons3}) or explicitly in Eq.(\ref{eq:2ndcoeff}) for multi-frequency methods, which is the Type III gate. As shown as the green dots in Fig~\ref{fig:t-err}, the infidelities from out-of-Lamb-Dicke errors for the Type III gate are significantly suppressed to $10^{-2}$ or even around $10^{-4}$ for some cases near the duration comparable to one-vibrational period. We include the second order constraint for the initial phase robustness in Appendix~{\ref{App: Initial phase robustness condition}} and observe much suppressed dependence of the initial phase as shown in Fig~\ref{fig:t-err}(b). The peak of the green curve at slightly below $2\mu s$ might originate from the resonance of some frequency components  to larger than second-order vibrational modes, since the constraints up to second order of Lamb-Dicke approximation are included.


	\begin{table}[htp!]
		\begin{tabular}{lcc}
			\hline
			& assumption & error \\ \hline
			Out-of-Lamb-Dicke     &            & $1.2\times 10^{-3}$\\ \hline
			Optical phase chirp   &     & $\sim 10^{-3}$ \\ \hline
			Pulse timing          &  0.1ns   & $\sim10^{-3}$ \\ \hline
			Amplitude fluctuation &  1\%       & $10^{-4}$ \\ \hline
			Radial mode coupling  &  $1^{\circ}$   & $\sim10^{-6}$ \\ \hline
			Photon scattering     &     &  $1.2\times 10^{-4}$\\ \hline
			Motional mode heating &  $\dot{\bar{n}}=100$   & $\sim 10^{-6}$ \\ \hline  
		\end{tabular}
		\caption{The error analysis for the Type III gate with the duration $\approx 0.79\mu s$. The details of the estimations are discussed in Appendix~\ref{App:Error budgets estimation}.}
		\label{tab:2_qubit_errors}
	\end{table}
	
	
	We further analyze other sources of errors and summarize them in Table~\ref{tab:2_qubit_errors} for the Type III gate with the duration $\approx 0.79~\mu$s. The overall error is in the order of $10^{-3}$ and it can be further suppressed in principle. The dominant error of $1.2 \times 10^{-3}$ comes from out-of-Lamb-Dicke approximation. It can be further suppressed by an order of magnitude if the Lamb-Dicke parameter $\eta_{m}=0.13$ is reduced to $\eta_{m}=0.07$. It can be seen that this gate is rather sensitive to optical phase chirp and pulse timing errors. The optical-phase-chirp error can be reduced by adding phase modulation that counters chirping. in principle. There have been many developed methods \cite{Shapira2018Robust,Webb2018Rsilient,Shapira2020-xe} to make the gate robust against the timing errors, which can supplement our gate scheme straightforwardly.

	\begin{figure}[htp!]
		\centering
		\includegraphics[width=1.0\columnwidth]{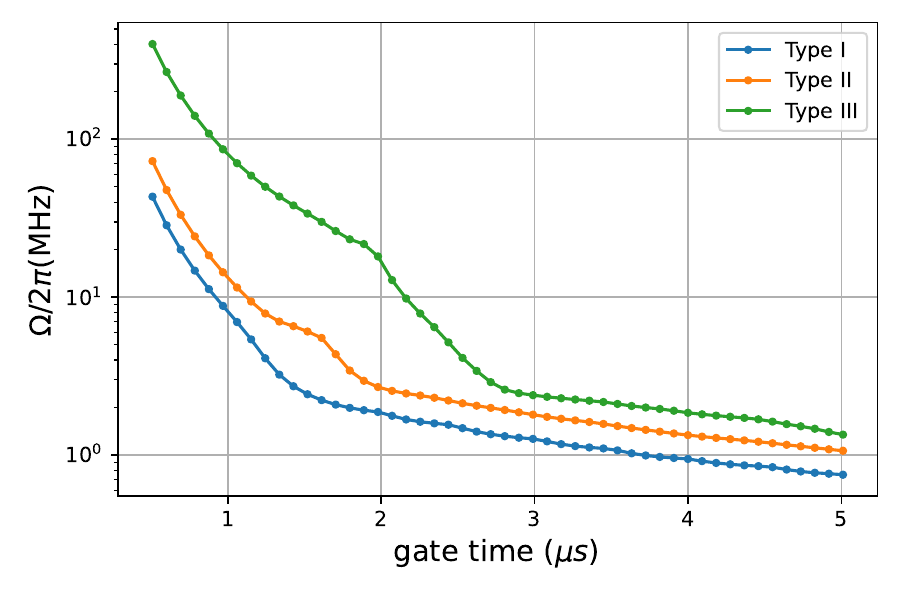}
		\caption{ The required Rabi frequency depending on the duration of two-qubit gate. The colors of the three curves corresponding to the same case as Fig.~\ref{fig:t-err}.}
		\label{fig:t-Omega}
	\end{figure}

	We also investigate the required Rabi-frequency with the gate time. For the 2-qubit gate, the waveform with the lowest Rabi-frequency can be calculated analytically applying the formalism described above as discussed in Ref.\cite{Shapira2020-xe}. The entangling phase $\Theta_{12}=\frac{1}{2}(\Theta_{m=0}-\Theta_{m=1})$ for a standard $\hat\sigma_{x}\hat\sigma_{x}$ gate of two qubits is $\frac{\pi}{4}$. We can analytically solve the quadratic Eq.(\ref{eq:Theta_m}) related to  $\Theta_{12}=\pi/4$ and find the largest eigenvalue and the corresponding eigenvector of its coefficient matrix $A_0 - A_1$, which leads to the solution with the lowest Rabi-frequency. However, in more complicated cases, for instance while taking into account the initial phase robustness and second order constraints, we can find solutions only by numerical methods that use the Python Scipy package with the Sequential Least Squares Programming method~\cite{2020SciPy-NMeth}. Here, to be consistent in all the results, we numerically minimize the Rabi frequency with the restriction of the quadratic constraints for all types of the gates, which is summarized in Fig.~\ref{fig:t-Omega}. 
	
	The Type I gates require smaller Rabi frequencies than the Type II and the Type III gates. The Type II  gates require about twice as much Rabi frequencies as the Type I gate, while the Type III gates demand an order of magnitude larger Rabi frequency at the expense of robustness against the initial phase fluctuation and inclusion of the second-order constraints. For the case of $^{171}\mathrm{Yb}^+$ ions with 355 nm laser, carrier Rabi-frequency about 1 MHz can be achieved with 1 mW power focused on 2 $\mu$m spots for each beams \cite{Lu2019-yg}. Therefore, it can be within experimental reach for the Type III gate with around 1 $\mu$s duration, which can be realized with 100 mW per beam.      
	
	\subsection{\label{subsec:three-qubit}Three-qubit gate}
	
\begin{figure}[htp!]
		\centering
		\includegraphics[width=1.0\columnwidth]{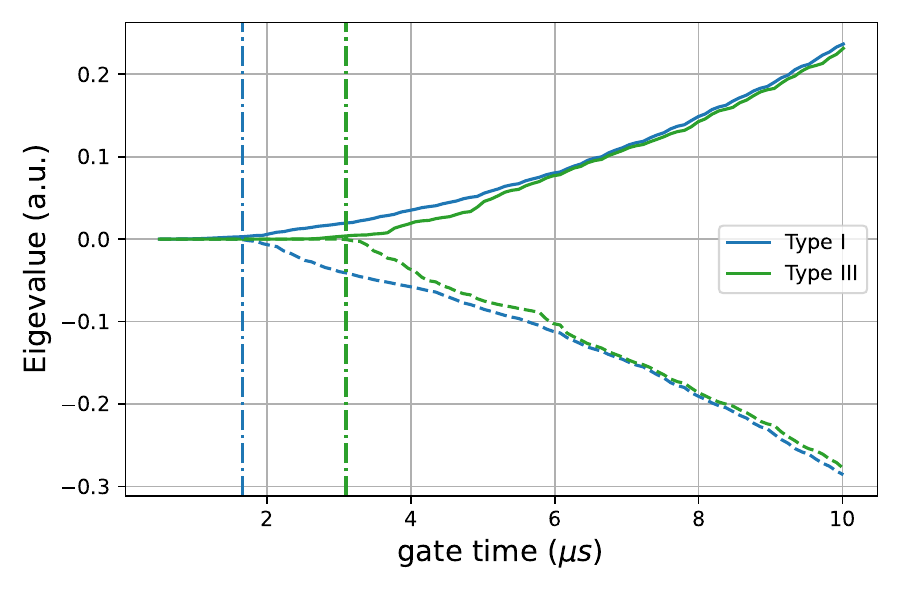}
		\caption{The evaluation of minimum gate time from eigenvalue methods, The dashed lines represent the largest negative eigenvalues while the solid lines correspond to the smallest positive eigen values, and the vertical lines corresponds to the lower boundary of the gate time, where matrix in Eq.({\ref{eq:Theta_3a}}) becomes semi-positive or semi-negative. As this eigenvalue method can only deal with one quadratic constraints, for the Type I gate, it shows the exact boundary. Since the eigenvalue method cannot be applied with more than one quadratic constraints. the eigenvalues for the Type II and the Type III are obtained without additional constraints for initial phase robustness. Therefore, the eigenvalues for the Type II gate is exactly same to those of the Type I gate.}
		\label{fig:3t-Omega}
	\end{figure}

	Different from two-qubit gate, there exists a minimum gate time in the global entangling gates for larger than three qubits. For the case of three qubits, we can evaluate the minimum gate time of the global gates from the existence of the solutions for Eq.(\ref{eq:Theta_m}) depending on the gate time. We can find the tight bound for the case of the Type-I gate. The phase requirements for the Type-I global-gate with three qubits, $\Theta_{12}=\Theta_{13}=\Theta_{23}=\frac{\pi}{4}$, can be rewritten as 
	\begin{eqnarray}
		\Theta_{m=1}-\Theta_{m=2}&=0 \label{eq:Theta_3a}\\
		\Theta_{m=0}-\Theta_{m=1}&=\frac{3\pi}{4}. \label{eq:Theta_3b}
	\end{eqnarray}
	The quadratic equation Eq.(\ref{eq:Theta_3a}) can be addressed by dividing the eigen-vectors of the matrix $A_1-A_2$ into two groups, one with positive eigenvalues the other with negative ones. As long as there exists both positive and negative eigenvalues we can always find a solution of $\vec{x}$~\cite{Shapira2020-xe}. Therefore, we can evaluate the lower bound of the gate time by solving the eigenvalue equation of the matrix $A_1-A_2$ depending on the gate time and finding when the matrix becomes semi-positive or semi-negative definite. The maximum and minimum eigenvalues of the matrix are numerically found and shown as the blue solid and dashed lines in Fig.~\ref{fig:3t-Omega}. The blue dashed-dot line at 1.66 $\mu$s shows the lower bound of the gate time.

	For the Type II gate that contains the initial phase robustness, we include additional two quadratic conditions shown in Eq.(\ref{Eq: initial phase robust condition3}) and (\ref{Eq: initial phase robust condition4}). Because of the additional quadratic constraints, we do not have a single equation that enables us to evaluate the minimum gate time by searching the zeros of eigenvalues of the corresponding matrix. Instead, we evaluate the minimum gate time by numerically finding solutions with all quadratic constraints. In this way, we find the minimum gate time of 1.71 $\mu$s that is close to the boundary by the eigenvalue method shown in the blue vertical line of Fig.~\ref{fig:3t-Omega}. 
	
	The Type III gate also contains the initial phase robustness and we cannot apply the eigenvalue method. However, we first neglect the additional quadratic constraints for the initial phase robustness and find the boundary, where the eigenvalues of the corresponding matrix of Eq.(\ref{eq:Theta_3a}). The maximum and minimum eigenvalues of the matrix for the Type III gate are shown as the green solid and dashed lines in Fig.~\ref{fig:3t-Omega}, where the green dashed-dot line at 3.10 $\mu$s shows the time when zeros of eigenvalues occurs. Similar to the case of the Type II gate, we numerically search solutions with all quadratic constraints and find the minimum gate time of 3.21 $\mu$s. Therefore, those boundaries found by the eigenvalue methods provide a decent estimation of the actual boundaries. 
	
	
	\subsection{\label{subsec:four-qubit}N-qubit gates (N$>$3)}
\begin{figure}[htp!]
		\includegraphics[width=1\columnwidth]{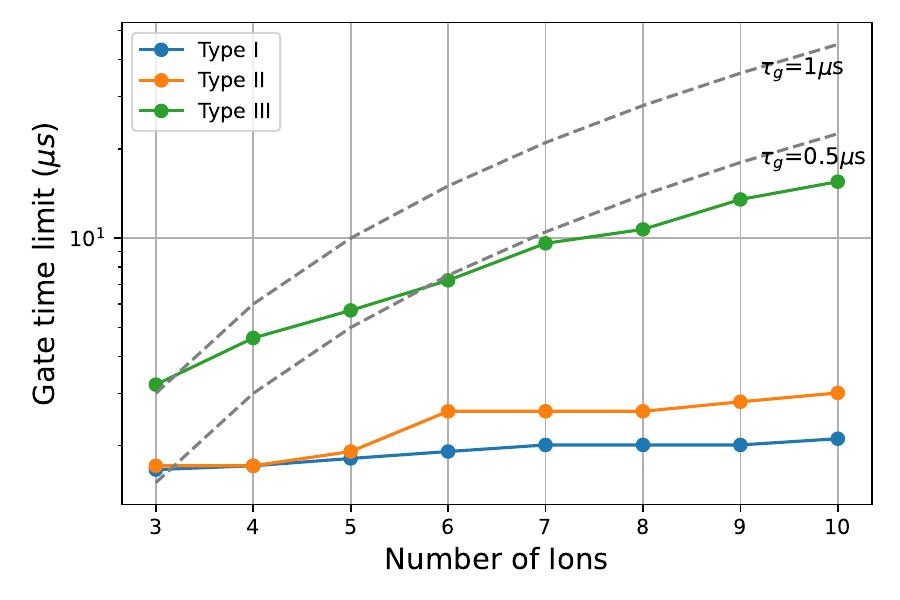}
		\caption{ Minimum gate times for N-qubit global-gates from 3 to 10 qubits, for all the Types. The colors for the Types are same to those in  Fig.~\ref{fig:t-err}. The grey curves represent the time used to achieve the same gate using a sequence of two-qubit gates with gate time $0.5\mu s$ and $1\mu s$ respectively.}
		\label{fig:N-qubit bound}
	\end{figure}

	The gate time for the N-qubit global gate (N$>$3) cannot be shorter than that of the three-qubit gate, since the constraints for the N-qubit gate comprise those of the three-qubit gate. The eigenvalue method for the evaluation of the minimum gate time cannot be used for more than three-qubit global-gates, we numerically search the minimum gate time by finding solutions of the gates with all constraints. In practice, we estimate the lower boundary of the gate time based on the existence of a numerical solutions within certain iteration steps (here set to $10^6$). The boundaries of gate time for all the Types from 3- to 10-qubit gates are shown in Fig.~\ref{fig:N-qubit bound}. For the Type I and the Type II gates, the durations increases from 1.66 $\mu$s and 1.71 $\mu$s to 2.11 $\mu$s and 3.01 $\mu$s, respectively, when the number of ions increase from 3 to 10. To our surprise, the gate times are increased a few times when we include the second order constraints from $3.21\mu$s to $15.51\mu$s.   
	
	An N-qubit gate can be decomposed into $\frac{N(N+1)}{2}$ of 2-qubit gates. In Fig.~\ref{fig:N-qubit bound}, the dashed grey lines represent the total gate times expected when the global gates are decomposed into multiple 2-qubit gates with $0.5~\mu s$ or $1~\mu s$ gate time. We find that even for Type III gate, the durations of the global gates are less than the total times of equivalent ultra-fast multiple of two-qubit-gates as the number of ions increase. We note that theoretically, one can achieve arbitrarily fast N-qubit gate with individual addressing of the qubits. 
	
	The error of the multi-qubit gates are difficult to exactly evaluate because of the limited dimension of Hilbert space that we are able to simulate. We expect the fidelities of N-qubit global gates would be bounded by those of two-qubit gates with the same duration. It would be possible that the long gate time with the second order constraints comes from the strict requirements of nullifying the effects of the second order terms and there exist N-qubit global gates with shorter duration by releasing the second order constraints. 
	
	We notice that the required Rabi frequency and the excited phonon number diverge at the gate time boundaries. Slowing down the gate by only hundreds of nanosecond, we can significantly reduce the required Rabi frequency and phonon excitation where we wouldn't expect a large out-of-Lamb-Dicke error, especially when the second order constraints are considered.

	
	\section{\label{sec:Conc}Conclusion and Outlook\protect\\}
	
	We have presented the schemes of realizing fast entangling gates on N-qubit registers without individual addressing. By applying the multi-frequency methods \cite{Shapira2020-xe}, we have found two-qubit gates below one motional period with only below $10^{-3}$ level of out-of-Lamb-Dicke error, which is considered as one of the major fundamental errors for gates at such speed. The same methods are also applied to N-qubit (N$>$2) systems, where we have found the speed limit of the global entangling gates. For three-qubit system a theoretical lower boundary is given, while for larger systems we estimate the limits based on numerical searching. The scaling of the gate time limit is found to be superior to achieving the equivalent operations with sequences of two-qubit gates. The gates can be insensitive to initial phase fluctuation which is considered difficult to stabilise in practice.
	
	We believe that, with the capability of controlling individual qubits, the speed of the gate can be further improved, but it may require more computational power to optimize larger number of control variables and more complex experimental setups\cite{Lu2019-yg}. We only include the robustness condition for the initial phase fluctuation. However, as indicated in Table~\ref{tab:2_qubit_errors}, the gates can be sensitive to other experiment parameters. There are many proposed methods with which we can further optimize our gate reducing its sensitivity to other parameters and other sources of errors \cite{Shapira2018Robust,Shapira2020-xe,Leung2018Robust, Milne2020Phase, Khaneja2005-aj, Bentley2020-ei}.
	
	\begin{acknowledgments}
		KZW and JFY are equally contributed to the work. The correspondence of the paper is to KZW, JNZ and KK. This work was supported by the National Key Research and Development Program of China under Grants No.\ 2016YFA0301900 and No.\ 2016YFA0301901, the National Natural Science Foundation of China Grants No.\ 92065205, and No.\ 11974200.
	\end{acknowledgments}

	\bibliography{apssamp}
	
	\clearpage
	\appendix
	
	\onecolumngrid
	\section{Derivation of the evolution operator\label{App: evol}}
	The Evolution operator in Eq.(\ref{Eq:Evol}) can be written as
	\begin{equation}
		\hat U(\tau)=\exp\left(\hat{A}_0+\hat{A}_1+\hat{A}_2+\hat{A}_{11}\right)
	\end{equation}
	where,
	\begin{eqnarray}
		\hat{A}_0&=&\frac{i}{\hbar}\int_{0}^{\tau_g}\hat{\mathcal{H}}_0(t)\mathrm{d}t =-i\Phi\sum_{j=1}^{N}\hat\sigma^j_x\\
		\hat{A}_1&=&\frac{i}{\hbar}\int_{0}^{\tau_g}\hat{\mathcal{H}}_1(t)\mathrm{d}t =\sum_{j=1}^{N}\sum_{m=0}^{N-1} \left(\alpha_{j,m}\hat{a}^{\dagger}_m-\alpha^*_{j,m}\hat{a}_m\right)\hat\sigma^j_x\\
		\hat{A}_2&=&\frac{i}{\hbar}\int_{0}^{\tau_g}\hat{\mathcal{H}}_2(t)\mathrm{d}t\nonumber\\&=&\sum_{j=1}^{N}\sum_{m_1,m_2=0}^{N-1}\left(\beta^{j}_{m_1,+m_2}\hat{a}^{\dagger}_{m_1}\hat{a}^{\dagger}_{m_2}+\beta^{j}_{m_1,-m_2}\hat{a}^{\dagger}_{m_1}\hat{a}_{m_2}-\beta^{j*}_{m_1,-m_2}\hat{a}_{m_1}\hat{a}^{\dagger}_{m_2}-\beta^{j*}_{m_1,+m_2}\hat{a}_{m_1}\hat{a}_{m_2}\right)\\
		\hat{A}_{11}&=&-\frac{1}{2\hbar^2}\int_{0}^{\tau_g}\mathrm{d}t_1\int_{0}^{t_1}\mathrm{d}t\left[\hat{\mathcal{H}}_1(t_1),\;\hat{\mathcal{H}}_1(t)\right]{}=-i\sum_{1\leq i< j\leq N}\Theta_{ij}\hat\sigma^{i}_x\hat\sigma^{j}_x
		\label{Eq:Evol_explicit}
	\end{eqnarray}
	
	For our convenience, we further define $\alpha_m$, $\Theta_m$ and $\beta_{m_1,\pm m_2}$ corresponding to the motional modes such that 
	\begin{eqnarray}
		\alpha_{j,m}&=&b_{j,m}\alpha_m,\\
		\beta^{j}_{m_1,\pm m_2}&=&b_{j,m_1}b_{j,m_2}\beta_{m_1,\pm m_2}\\
		\Theta_{ij}&=&\sum_{m}b_{i,m}b_{j,m}\Theta_m. \label{eq:alphaTheta}
	\end{eqnarray}
	The explicit expression of $\Phi$, $\alpha_m$, $\beta_{m_1,\pm m_2}$ and $\Theta_{m}$ are written as
	\begin{eqnarray}
		\Phi(\tau)&=&\int_{0}^{\tau}\Omega(t)\cos(\mu t+\phi(t))\mathrm{d} t\\
		\alpha_{m}(\tau)&=&i\eta_m\int_{0}^{\tau}\Omega(t)\sin(\mu t+\phi(t))e^{i\nu_m t} \mathrm{d} t\\
		\beta_{m_1,\pm m_2}(\tau)&=&\frac{i\eta_{m_1}\eta_{m_2}}{2}\int_{0}^{\tau}\Omega(t)\cos(\mu t+\phi(t))e^{i(\nu_{m_1}\pm\nu_{m_2}) t} \mathrm{d} t\\
		\Theta_{m}(\tau)&=&-2\eta_{m}^2\int_{0}^{\tau}\int_{0}^{t_1}\Omega(t_1)\Omega(t_2)\sin(\mu t_1+\phi(t_1))\sin(\mu t_2+\phi(t_2))\sin(\nu_m(t_1-t_2))\mathrm{d} t_1\mathrm{d} t_2
	\end{eqnarray}
	
	\onecolumngrid
	\section{Matrices expression\label{App: Matrices expression}}
	
	The expressions of the entries in matrices $\vec{a}_m$, $\beta_{m_1, \pm m_2}$ and $\tilde{\theta}_m$ are given as the following.
	\begin{equation}
		(\vec{a}_m)_j=\left\{
		\begin{aligned}
			&\int_{0}^{\tau_g}\sin(\omega_j t)e^{i\nu_mt}\mathrm{d} t & 1\leq j\leq n_e\\
			&\int_{0}^{\tau_g}\sin(\omega_{-(j-n_e)} t)e^{i\nu_mt}\mathrm{d} t & n_e+1\leq j\leq 2n_e\\
			&\int_{0}^{\tau_g}\cos(\omega_{(j-2n_e)} t)e^{i\nu_mt}\mathrm{d} t & 2n_e+1\leq j\leq 3n_e\\
			&\int_{0}^{\tau_g}\cos(\omega_{-(j-3n_e)} t)e^{i\nu_mt}\mathrm{d} t & 3n_e+1\leq j\leq 4n_e
		\end{aligned}
		\right., \quad m=0,1,...,N-1
	\end{equation}
	\begin{equation}
		(\theta_m)_{ij}=\left\{
		\begin{aligned}
			&-2\eta_{m}^2\int_{0}^{\tau_g}\int_{0}^{t_1}\sin(i\omega t_1)\sin(j\omega t_2)&\\
			&\times\sin(\nu_m(t_1-t_2))\mathrm{d} t_1\mathrm{d} t_2 & 1\leq i\leq n_e,\;1\leq j\leq n_e\\
			&2\eta_{m}^2\int_{0}^{\tau_g}\int_{0}^{t_1}\sin(i\omega t_1)\sin((j-n_e)\omega t_2)&\\
			&\times\sin(\nu_m(t_1-t_2))\mathrm{d} t_1\mathrm{d} t_2 & 1\leq i\leq n_e,\;n_e+1\leq j\leq 2n_e\\
			&-2\eta_{m}^2\int_{0}^{\tau_g}\int_{0}^{t_1}\sin(i\omega t_1)\cos((j-2n_e)\omega t_2)&\\
			&\times\sin(\nu_m(t_1-t_2))\mathrm{d} t_1\mathrm{d} t_2 & 1\leq i\leq n_e,\;2n_e+1\leq j\leq 3n_e\\
			&-2\eta_{m}^2\int_{0}^{\tau_g}\int_{0}^{t_1}\sin(i\omega t_1)\cos((j-3n_e)\omega t_2)&\\
			&\times\sin(\nu_m(t_1-t_2))\mathrm{d} t_1\mathrm{d} t_2 & 1\leq i\leq n_e,\;3n_e+1\leq j\leq 4n_e\\
			&2\eta_{m}^2\int_{0}^{\tau_g}\int_{0}^{t_1}\sin((i-n_e)\omega t_1)\sin(j\omega t_2)&\\
			&\times\sin(\nu_m(t_1-t_2))\mathrm{d} t_1\mathrm{d} t_2 & n_e+1\leq i\leq 2n_e,\;1\leq j\leq n_e\\
			&-2\eta_{m}^2\int_{0}^{\tau_g}\int_{0}^{t_1}\sin((i-n_e)\omega t_1)\sin((j-n_e)\omega t_2)&\\
			&\times\sin(\nu_m(t_1-t_2))\mathrm{d} t_1\mathrm{d} t_2 & n_e+1\leq i\leq 2n_e,\;n_e+1\leq j\leq 2n_e\\
			&2\eta_{m}^2\int_{0}^{\tau_g}\int_{0}^{t_1}\sin((i-n_e)\omega t_1)\cos((j-2n_e)\omega t_2)&\\
			&\times\sin(\nu_m(t_1-t_2))\mathrm{d} t_1\mathrm{d} t_2 & n_e+1\leq i\leq 2n_e,\;2n_e+1\leq j\leq 3n_e\\
			&2\eta_{m}^2\int_{0}^{\tau_g}\int_{0}^{t_1}\sin((i-n_e)\omega t_1)\cos((j-3n_e)\omega t_2)&\\
			&\times\sin(\nu_m(t_1-t_2))\mathrm{d} t_1\mathrm{d} t_2 & n_e+1\leq i\leq 2n_e,\;3n_e+1\leq j\leq 4n_e\\
			&-2\eta_{m}^2\int_{0}^{\tau_g}\int_{0}^{t_1}\cos((i-3n_e)\omega t_1)\sin(j\omega t_2)&\\
			&\times\sin(\nu_m(t_1-t_2))\mathrm{d} t_1\mathrm{d} t_2 & 2n_e+1\leq i\leq 3n_e,\;1\leq j\leq n_e\\
			&2\eta_{m}^2\int_{0}^{\tau_g}\int_{0}^{t_1}\cos((i-3n_e)\omega t_1)\sin((j-n_e)\omega t_2)&\\
			&\times\sin(\nu_m(t_1-t_2))\mathrm{d} t_1\mathrm{d} t_2 & 2n_e+1\leq i\leq 3n_e,\;n_e+1\leq j\leq 2n_e\\
			&-2\eta_{m}^2\int_{0}^{\tau_g}\int_{0}^{t_1}\cos((i-3n_e)\omega t_1)\cos((j-2n_e)\omega t_2)&\\
			&\times\sin(\nu_m(t_1-t_2))\mathrm{d} t_1\mathrm{d} t_2 & 2n_e+1\leq i\leq 3n_e,\;2n_e+1\leq j\leq 3n_e\\
			&-2\eta_{m}^2\int_{0}^{\tau_g}\int_{0}^{t_1}\cos((i-3n_e)\omega t_1)\cos((j-3n_e)\omega t_2)&\\
			&\times\sin(\nu_m(t_1-t_2))\mathrm{d} t_1\mathrm{d} t_2 & 2n_e+1\leq i\leq 3n_e,\;3n_e+1\leq j\leq 4n_e\\    &-2\eta_{m}^2\int_{0}^{\tau_g}\int_{0}^{t_1}\cos((i-4n_e)\omega t_1)\sin(j\omega t_2)&\\
			&\times\sin(\nu_m(t_1-t_2))\mathrm{d} t_1\mathrm{d} t_2 & 3n_e+1\leq i\leq 4n_e,\;1\leq j\leq n_e\\
			&2\eta_{m}^2\int_{0}^{\tau_g}\int_{0}^{t_1}\cos((i-4n_e)\omega t_1)\sin((j-n_e)\omega t_2)&\\
			&\times\sin(\nu_m(t_1-t_2))\mathrm{d} t_1\mathrm{d} t_2 & 3n_e+1\leq i\leq 4n_e,\;n_e+1\leq j\leq 2n_e\\
			&-2\eta_{m}^2\int_{0}^{\tau_g}\int_{0}^{t_1}\cos((i-4n_e)\omega t_1)\cos((j-2n_e)\omega t_2)&\\
			&\times\sin(\nu_m(t_1-t_2))\mathrm{d} t_1\mathrm{d} t_2 & 3n_e+1\leq i\leq 4n_e,\;2n_e+1\leq j\leq 3n_e\\
			&-2\eta_{m}^2\int_{0}^{\tau_g}\int_{0}^{t_1}\cos((i-4n_e)\omega t_1)\cos((j-3n_e)\omega t_2)&\\
			&\times\sin(\nu_m(t_1-t_2))\mathrm{d} t_1\mathrm{d} t_2 & 3n_e+1\leq i\leq 4n_e,\;3n_e+1\leq j\leq 4n_e
		\end{aligned}
		\right.
	\end{equation}
	and $\tilde{\theta}_m=\frac{1}{2}\left(\theta_m+\theta_m^{T}\right)$
	
	\begin{equation}
		(\vec{b}_{m_1,\pm m_2})_j=\left\{
		\begin{aligned}
			&\int_{0}^{\tau_g}\sin(\omega_j t)e^{i(\nu_{m_1}\pm\nu_{m_2})t}\mathrm{d} t & 1\leq j\leq n_e\\
			&\int_{0}^{\tau_g}\sin(\omega_{-(j-n_e)} t)e^{i(\nu_{m_1}\pm\nu_{m_2})t}\mathrm{d} t & n_e+1\leq j\leq 2n_e\\
			&\int_{0}^{\tau_g}\cos(\omega_{(j-2n_e)} t)e^{i(\nu_{m_1}\pm\nu_{m_2})t}\mathrm{d} t & 2n_e+1\leq j\leq 3n_e\\
			&\int_{0}^{\tau_g}\cos(\omega_{-(j-3n_e)} t)e^{i(\nu_{m_1}\pm\nu_{m_2})t}\mathrm{d} t & 3n_e+1\leq j\leq 4n_e
		\end{aligned}
		\right., \quad m_1, m_2=0,1,...,N-1
	\end{equation}
	\onecolumngrid
	\section{Initial phase robustness condition\label{App: Initial phase robustness condition}}

	Expressing the dependency of the Hamiltonian to the initial phase $\phi$ explicitly, we have
	\begin{equation}
		\begin{aligned}
			\hat {\mathcal H}_0=&\sum_{j=1}^{N}\hbar\Omega\hat{\sigma}_{x}^{j}\sum_{n\neq 0}\left[c_{n}\cos(\omega_n t+\phi)-s_{n}\sin(\omega_n t+\phi)\right]\\
			=&\sum_{j=1}^{N}\hbar\Omega\hat{\sigma}_{x}^{j}\sum_{n\neq 0}\left[(c_{n}\cos(\omega_n t)-s_{n}\sin(\omega_n t))\cos\phi
			+(-s_{n}\cos(\omega_n t)-c_{n}\sin(\omega_n t))\sin\phi\right]\\
			\hat{\mathcal H}_1=&\sum_{j=1}^{N}\hbar\Omega\hat\sigma_{x}^j\sum_{m=0}^{N-1}\eta_{j,m}(\hat{a}^{\dagger}_m e^{i\nu_{m} t}+\hat{a}_{m}e^{-i\nu_{m} t})\sum_{n\neq 0}
			\left[-c_n\sin(\omega_n t+\phi)-s_n\cos(\omega_n t+\phi)\right]\\
			=&\sum_{j=1}^{N}\hbar\Omega\hat\sigma_{x}^j\sum_{m=0}^{N-1}\eta_{j,m}(\hat{a}^{\dagger}_m e^{i\nu_{m} t}+\hat{a}_{m}e^{-i\nu_{m} t})\\
			&\cdot\sum_{n\neq 0}\left[(-c_{n}\sin(\omega_n t)-s_{n}\cos(\omega_n t))\cos\phi
			+(s_{n}\sin(\omega_n t)-c_{n}\cos(\omega_n t))\sin\phi\right]\\
			\hat{\mathcal H}_2=&\sum_{j=1}^{N}\hbar\Omega\hat\sigma_{x}^j\left[\sum_{m=0}^{N-1}\eta_{j,m}(\hat{a}^{\dagger}_m e^{i\nu_{m} t}+\hat{a}_{m}e^{-i\nu_{m} t})\right]^2\sum_{n\neq 0}
			\left[c_n\cos(\omega_n t+\phi)-s_n\sin(\omega_n t+\phi)\right]\\
			=&\sum_{j=1}^{N}\hbar\Omega\hat\sigma_{x}^j\left[\sum_{m=0}^{N-1}\eta_{j,m}(\hat{a}^{\dagger}_m e^{i\nu_{m} t}+\hat{a}_{m}e^{-i\nu_{m} t})\right]^2\\
			&\cdot\sum_{n\neq 0}\left[(c_{n}\cos(\omega_n t)-s_{n}\sin(\omega_n t))\cos\phi
			+(-s_{n}\cos(\omega_n t)-c_{n}\sin(\omega_n t))\sin\phi\right]
		\end{aligned}
	\end{equation}
	Having this Hamiltonian, we can also write the parameters in the evolution operator as functions of the initial phase $\phi$. Define $\vec{\Omega}'=\Omega\begin{bmatrix}
		\vec{-s}\\
		\vec{c}
	\end{bmatrix}$, we have
	
	\begin{equation}
		\begin{aligned}
			\alpha_m(\phi)&=\vec{a}_{m}^{T}\vec{\Omega}\cos\phi+\vec{a}_{m}^{T}\vec{\Omega}'\sin\phi\\
			\beta_{m_1, \pm m_2}(\phi)&=\vec{b}_{m_1, \pm m_2}^{T}\vec{\Omega}\cos\phi+\vec{b}_{m_1, \pm m_2}^{T}\vec{\Omega}'\sin\phi\\
			\Theta_{m}(\phi)&=\vec{\Omega}^{T}\tilde{\theta}_m\vec{\Omega}\cos^2\phi+\vec{\Omega}'^{T}\tilde{\theta}_m\vec{\Omega}'\sin^2\phi
			-2\vec{\Omega}^{T}\tilde{\theta}_m\vec{\Omega}'\sin\phi\cos\phi
		\end{aligned}
	\end{equation}
	Therefore, the following conditions are necessary and sufficient for an initial phase robust gate.
	\begin{eqnarray}
		\vec{a}_{m}^{T}\vec{\Omega}=\vec{a}_{m}^{T}\vec{\Omega}'=0\\
		\vec{b}_{m_1, \pm m_2}^{T}\vec{\Omega}=\vec{b}_{m_1, \pm m_2}^{T}\vec{\Omega}'=0\\
		\vec{\Omega}^{T}\tilde{\theta}_m\vec{\Omega}=\vec{\Omega}'^{T}\tilde{\theta}_m\vec{\Omega}=\Theta_{m}\label{Eq: initial phase robust condition3}\\
		\vec{\Omega}^{T}\tilde{\theta}_m\vec{\Omega}'=0\label{Eq: initial phase robust condition4}
	\end{eqnarray}

	\onecolumngrid
	\section{2-qubit gate example \label{App: 2-qubit example}}
	
	\begin{figure*}[htp!]
		\begin{center}
			\includegraphics[width=1\textwidth]{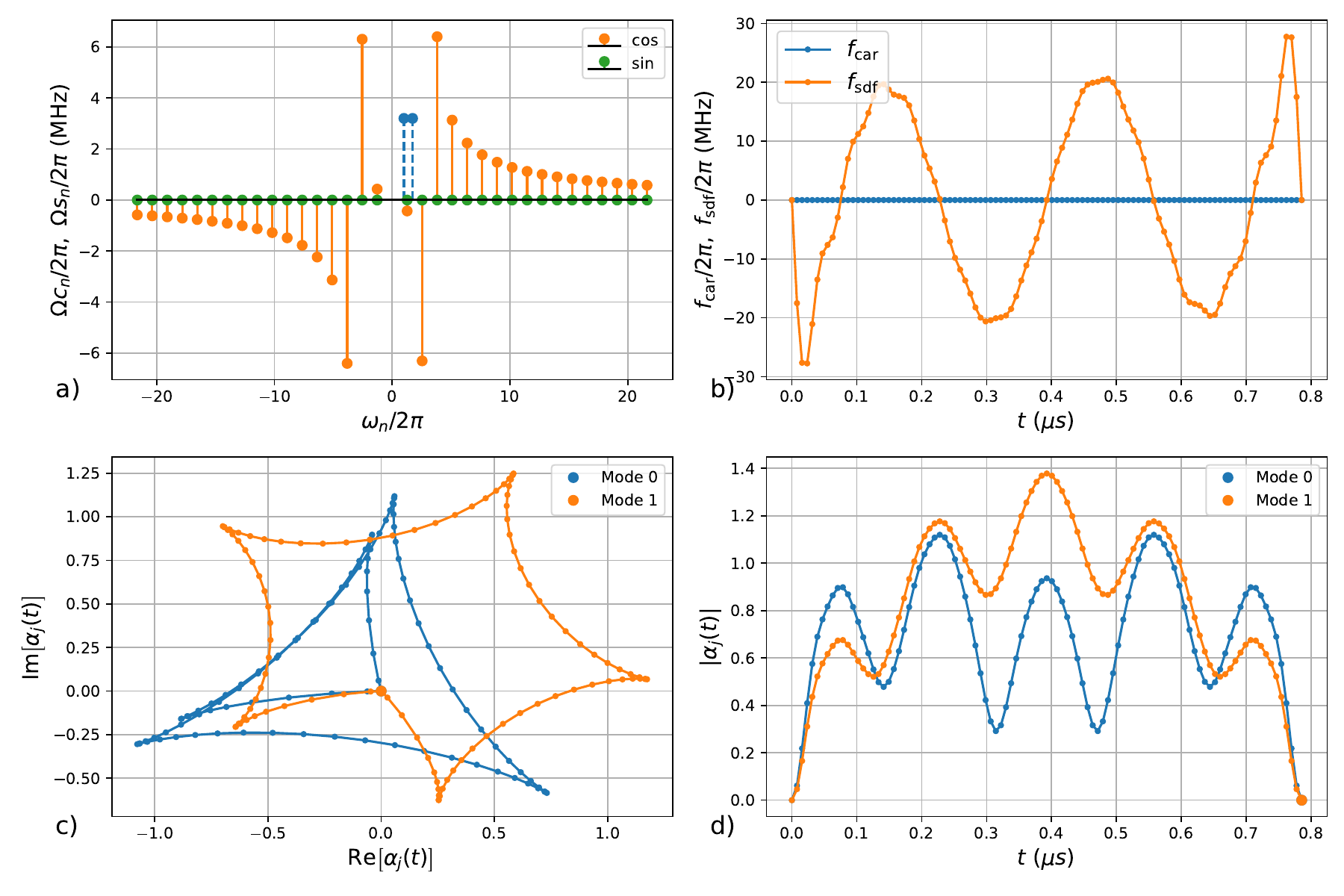}
		\end{center}
		\caption{Example of Type I fast two-qubit gate with gate time at approximately $0.79\mu s$. (a) Rabi frequencies required for each frequency components. The blue dashed lines corresponds to the motional frequency. (b) The time dependent coefficient $f_{car}$ and $f_{sdf}$ given in Eq.(\ref{Eq:fcar}) and Eq.(\ref{Eq:fsdf}). (c) The trajectories of different motional modes in the phase space. (d) The absolute value of the displacements from the origin in the phase space.
		}
		\label{fig:2-qubit ex1}
	\end{figure*}
	
	\begin{figure*}[htp!]
		\begin{center}
			\includegraphics[width=1\textwidth]{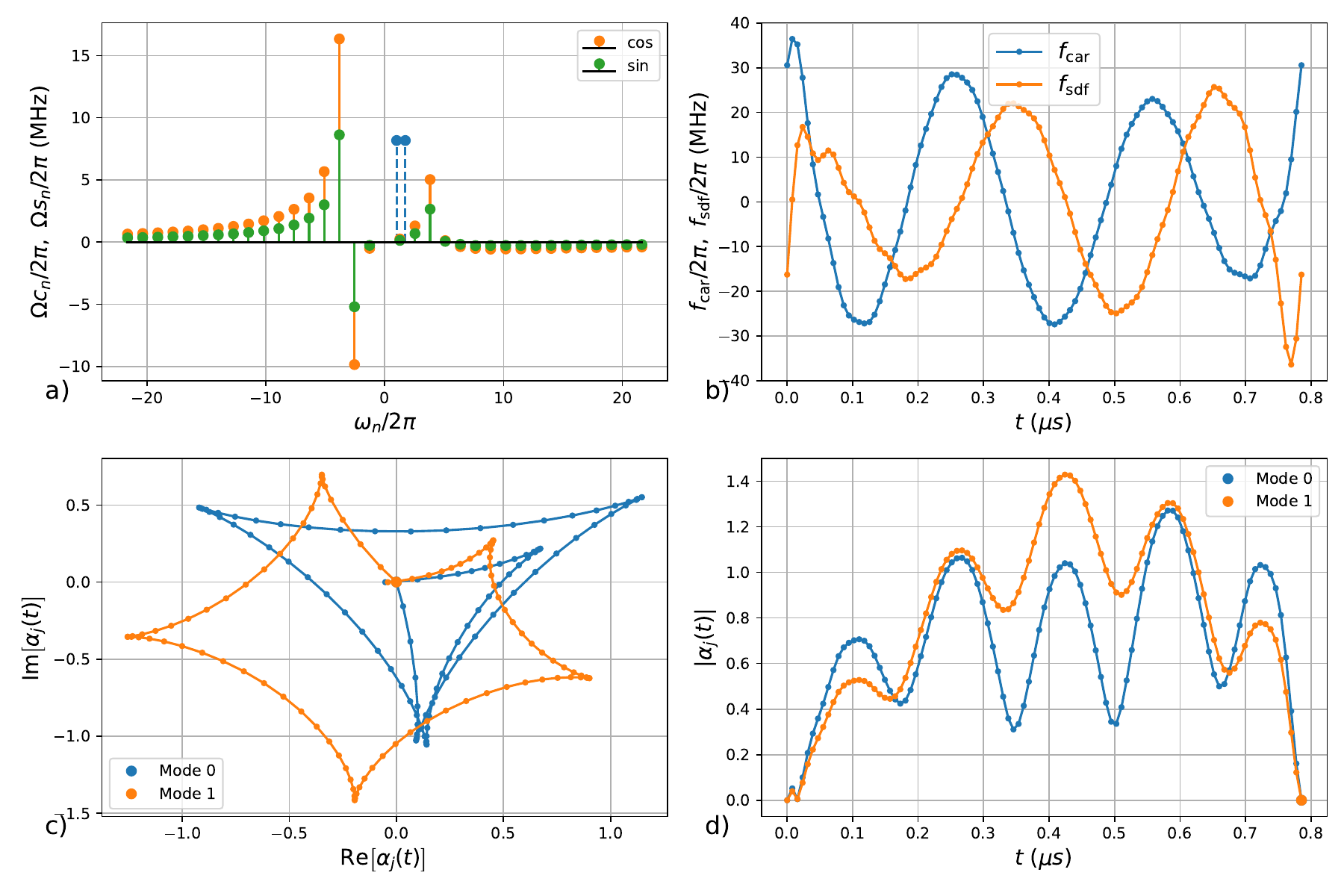}
		\end{center}
		\caption{Example of Type II fast two-qubit gate with gate time at approximately $0.79\mu s$. (a) Rabi frequencies required for each frequency components. The blue dashed lines corresponds to the motional frequency. (b) The time dependent coefficient $f_{car}$ and $f_{sdf}$ given in Eq.(\ref{Eq:fcar}) and Eq.(\ref{Eq:fsdf}). (c) The trajectories of different motional modes in the phase space. (d) The absolute value of the displacements from the origin in the phase space.
		}
		\label{fig:2-qubit ex3}
	\end{figure*}
	
	\begin{figure*}[htp!]
		\begin{center}
			\includegraphics[width=1\textwidth]{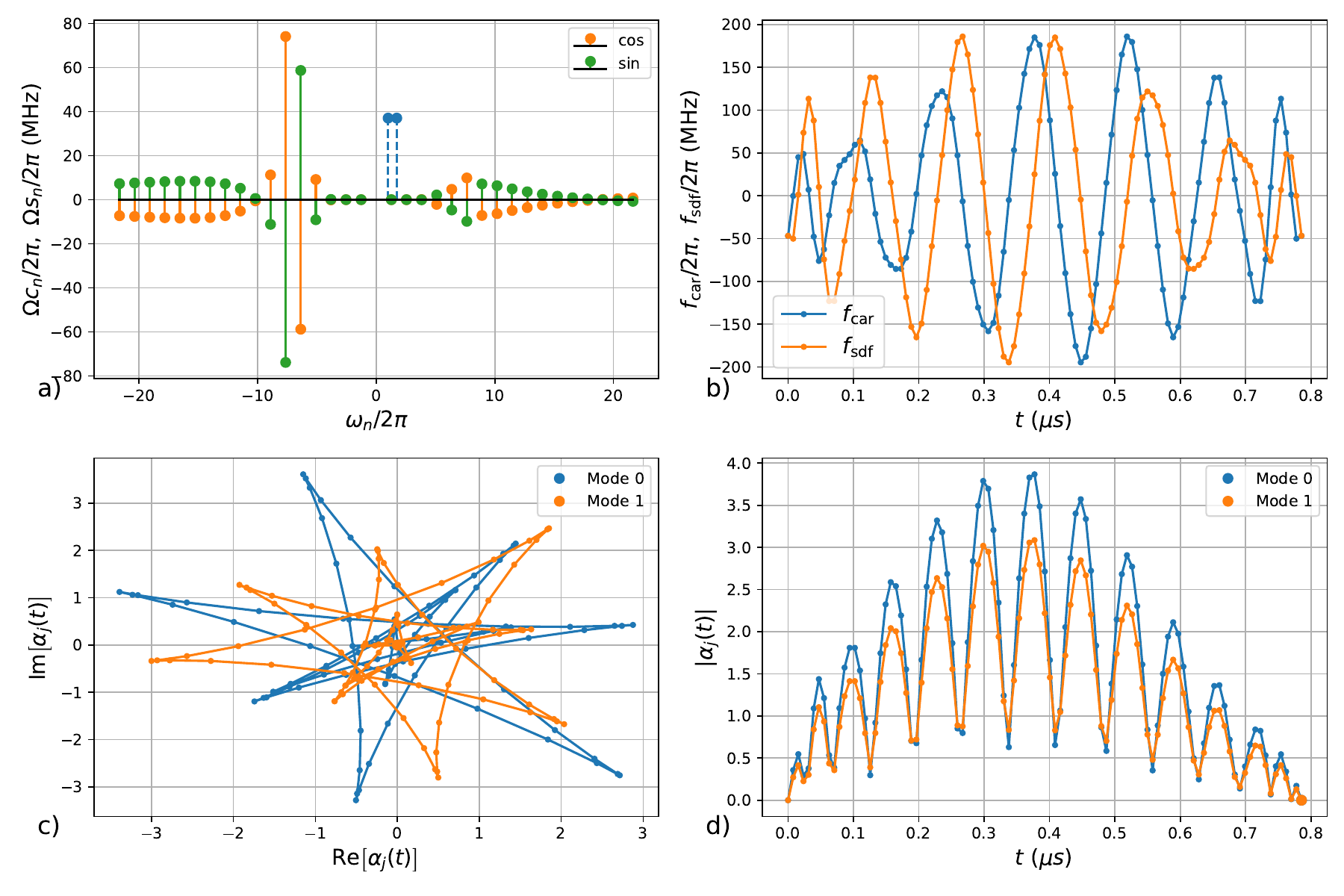}
		\end{center}
		\caption{Example of Type III fast two-qubit gate with gate time at approximately $0.79\mu s$. (a) Rabi frequencies required for each frequency components. The blue dashed lines corresponds to the motional frequency. (b) The time dependent coefficient $f_{car}$ and $f_{sdf}$ given in Eq.(\ref{Eq:fcar}) and Eq.(\ref{Eq:fsdf}). (c) The trajectories of different motional modes in the phase space. (d) The absolute value of the displacements from the origin in the phase space.
		}
		\label{fig:2-qubit ex4}
	\end{figure*}
	
	\onecolumngrid
	\section{Error budgets estimation \label{App:Error budgets estimation}}
	
	Here, we estimate some potential sources of error, assuming implement the gate with multi-frequency components.
	
	\subsection{Out-of-Lamb-Dicke errors}
	The out-of-Lamb-Dicke errors are estimated by using the time evolution of the original Hamiltonian (\ref{Eq:Hamiltonian_full}) without approximation and averaging the infidelities depending on the initial phases shown in Fig.~\ref{fig:t-err}(b).
	
	\subsection{Optical phase chirp}
	The chirping error is estimated based on the method given in Ref.\cite{Schafer_undated-tq}. Assuming we have a square pulse with duration $\tau_g$ given by
	\begin{equation}
		f(t)=\Omega(t)\cos(\omega t+\phi)
	\end{equation}
	where
	\begin{equation}
		\Omega(t)=\left\{
		\begin{aligned}
			&\Omega_0&0\leq t\leq \tau_g\\
			&0& \text{Otherwise}
		\end{aligned}
		\right.
	\end{equation}
	
	Due to rapid turning on and off the pulse, there will be a phase chirp at the start and the end of the pulse. We represent this phase chirp in the following form.
	
	\begin{equation}
		\begin{aligned}{}
			f_{ch}(t)=&\Omega(t)\cos(\omega t+\phi+\phi_{ch}(t))\\
			=&\Omega(t)\left[\cos(\omega t+\phi)\cos\phi_{ch}(t)-\sin(\omega t+\phi)\sin\phi_{ch}(t)\right]\\
			\approx & \Omega(t)\left[\cos(\omega t+\phi)-\sin(\omega t+\phi)\phi_{ch}(t)\right]\\
			=&\Omega(t)\cos(\omega t+\phi)+\Omega_{ch}(t)\sin(\omega t+\phi)
		\end{aligned}
	\end{equation}
	where $\Omega_{ch}(t)=-\phi_{ch}(t)\Omega(t)$. Here, we assume
	\begin{equation}
		\Omega_{ch}(t)=\left\{
		\begin{aligned}
			&A_{ch}(1-\frac{t}{t_{ch}})\Omega_0&0\leq t \leq t_{ch}\\
			&A_{ch}(1+\frac{t-\tau_{ch}}{t_{ch}})\Omega_0&\tau_g-t_{ch}\leq t\leq \tau_{ch}\\
			&0&\text{Otherwise}
		\end{aligned}
		\right.
	\end{equation}
	We simulate the Hamiltonian under the Lamb-Dicke approximation with the waveform including the chirping effect, assuming $A_{ch}=0.03$ and $t_{ch}=5$ns, which gives an $\sim10^{-3}$ error.
	
	\subsection{Pulse timing and Amplitude fluctuation}
	
	For the timing and amplitude fluctuation, we assume the fluctuation for each of the parameters following a Gaussian distribution and simulate the gate infidelity using the Hamiltonian under the Lamb-Dicke approximation. Assuming the standard deviation of the Gaussian distribution for timing and Rabi frequency are $0.1$ns and $1\%$ respectively, we obtain an error of $10^{-3}$ and $10^{-4}$.
	
	\subsection{Radial mode coupling}
	
	We estimate the error of radial mode coupling by assuming the effective wave-vector of the two counter-propagating Raman transition beams are slight tilted from the axial direction of the trap. We then simulate the radial mode Hamiltonian which gives an $10^{-6}$ level infidelity assuming $1^\circ$ of misalignment and $2\pi\times3$MHz radial trap frequency.
	
	\subsection{Scattering error}
	We estimate the scattering error with the example of $^{171}{\rm Yb}^{+}$ ion with 355 nm laser beams. We consider the two beams of the Raman transition are all set to $\sigma^{+}$ polarization. Therefore, the Raman coupling rate, Raman scattering rate and Rayleigh dephasing rate can be calculated based on the expression given in Ref.\cite{Ballance_undated-qm}, we have
	\begin{eqnarray}
		\Omega_{Raman}&=&\frac{g^2}{2}\left(\frac{1}{3\Delta_{1/2}}-\frac{1}{3\Delta_{3/2}}\right)\\
		\Gamma_{Raman}&=&\frac{g^2\gamma}{4}\left(\frac{1}{9\Delta_{1/2}^2}+\frac{1}{9\Delta_{3/2}^2}\right)\\
		\Gamma_{el}&=&\frac{g^2\gamma}{4}\frac{1}{9\Delta_{3/2}^2}
	\end{eqnarray}
	where $g$ is the Rabi frequency for transition between S and P level, $\gamma$ is the spontaneous decay rate from the P levels, and $\gamma\approx 2\pi\times 20$ MHz. The detuning of $355$nm laser from $\mathrm{P}_{1/2}$ and $\mathrm{P}_{3/2}$ level $\Delta_{1/2}$ and $\Delta_{3/2}$ are $2\pi\times33$ THz and $-2\pi\times66$ THz, respectively and $\Omega_{Raman}\approx2\pi \times 141$ MHz for $\tau_g \approx$0.79 $\mu$s gate time. We can get $\Gamma_{Raman}\approx75~s^{-1}$ and $\Gamma_{el}\approx15~s^{-1}$. The error caused by these two effects can be estimated by $\epsilon_{Raman}= 2\Gamma_{Raman}\tau_g\approx1.2\times10^{-4}$ and $\epsilon_{el}=\frac{1}{2}\Gamma_{el}\tau_g\approx6\times10^{-6}$.
	\subsection{Motional heating}
	We can introduce the motional heating effect by introducing the Lindblad operator $L_{-}=\sqrt{\dot{\bar{n}}}\hat{a}$ and $L_{+}=\sqrt{\dot{\bar{n}}}\hat{a}^{\dagger}$. However, the simulation of master equation is difficult given the large phonon number needs to be included in the simulation. Therefore, we apply Eq.(65) in Ref.\cite{Sutherland2022-tt} to estimate the heating error. Applying that formula to our example a $100$ phonon per second heating rate will cause an error only at $10^{-6}$ level.

\end{document}